\documentstyle[aps,manuscript,psfig]{revtex}
\begin{document}
\draft
\renewcommand{\thefootnote}{\fnsymbol{footnote}}
\begin{title}
{\bf Power-Law Slip Profile of the Moving Contact Line in Two-Phase
    Immiscible Flows}
\end{title}
\author{Tiezheng Qian,$^1$ Xiao-Ping Wang,$^1$ and Ping Sheng$^2$}
\address{$^1$Department of Mathematics,
Hong Kong University of Science and Technology,\\
Clear Water Bay, Kowloon, Hong Kong, China\\
$^2$Department of Physics and Institute of Nano Science and Technology,\\
Hong Kong University of Science and Technology,\\ 
Clear Water Bay, Kowloon, Hong Kong, China} 
\maketitle
\begin{abstract}

Large scale molecular dynamics (MD) simulations on two-phase 
immiscible flows show that associated with the moving contact line, 
there is a very large $1/x$ partial-slip region where $x$ denotes 
the distance from the contact line. This power-law partial-slip region 
is verified in large-scale adaptive continuum simulations 
based on a local, continuum hydrodynamic formulation, 
which has proved successful in reproducing MD results at the nanoscale. 
Both MD and continuum simulations indicate the existence of 
a universal slip profile in the Stokes-flow regime, well
described by $v^{slip}(x)/V_w=1/(1+{x}/{al_s})$, where $v^{slip}$
is the slip velocity, $V_w$ the speed of moving wall, $l_s$ the 
slip length, and $a$ is a numerical constant.
Implications for the contact-line dissipation are discussed.

\end{abstract}
\narrowtext

Boundary condition that specifies the flow of fluid over 
a solid surface is a cornerstone of hydrodynamics. 
The no-slip condition, i.e., zero relative velocity between 
the fluid and solid at the interface, has been the paradigm 
in most of the hydrodynamics literature \cite{batchelor}.
In molecular dynamics (MD) simulations, however, a small amount of 
relative slip between the fluid and the solid surface is generally 
detected at high flow rate \cite{nbc}. Such slip can be accounted for 
by the Navier boundary condition (NBC), whereby the slip velocity is 
proportional to the tangential viscous stress \cite{nbc,ckb}.
The proportionality constant between the slip velocity $v^{slip}$ and 
the shear rate is denoted the slip length $l_s$, 
usually ranging from one to a few nanometers (from MD simulations). 
As the amount of slip is extremely small in subsonic flow rates, 
the NBC is practically indistinguishable from the no-slip condition 
in most situations. In contrast, for immiscible flows 
the MD simulations have shown near-complete slip in the vicinity 
of the moving contact line (MCL), defined as the intersection of 
fluid-fluid interface with the solid wall \cite{koplik,robbins,qws}.
An intriguing question ensues: In a mesoscopic or macroscopic system, 
what is the slip profile which consistently interpolates 
between the near complete slip at the MCL and the no-slip boundary condition 
that must hold at regions far away \cite{dussan,degennes}? 
Recent evidences have shown the slip profile obtained from 
nanoscale MD simulations to be accountable by the generalized 
Navier boundary condition (GNBC), in which the slip velocity 
is proportional to the total tangential stress --- 
the sum of the viscous stress and the uncompensated Young stress; 
the latter arises from the deviation of the fluid-fluid interface 
from its static configuration \cite{qws}.
Here we show through large-scale MD simulations and 
continuum hydrodynamic calculations, that there exists 
a power-law partial-slip region extending to hundreds of nanometers 
or even more, contrary to the usual expectation of 
a nanometer-scale slip region in the vicinity of the MCL.  
The existence of this large partial-slip region modifies 
the conventional picture that significant NBC slipping occurs 
only under high flow/shear rate. Instead, the power-law slip region 
is associated with the universal slip profile of the MCL, 
even at low flow rates.

MD simulations have been carried out for increasingly larger systems 
of immiscible Couette flow (Fig. 1). Two immiscible fluids were confined 
between two parallel walls in the $xy$ plane, 
with the fluid-solid boundaries defined by $z=0$, $H$. 
Periodic boundary conditions were imposed along the $x$ and $y$ directions.
Interaction between fluid molecules separated by a distance $r$ 
was modeled by a modified Lennard-Jones (LJ) potential 
$U_{ff}=4\epsilon\left[\left(\sigma/r\right)^{12}-
\delta_{ff}\left(\sigma/r\right)^6\right]$, where 
$\delta_{ff}=1$ for like molecules and $\delta_{ff}=-1$ for
molecules of different species.
The average number density for the fluids was set at $\rho=0.81/\sigma^3$. 
The temperature was controlled at $2.8\epsilon/k_B$. 
Each wall was constructed by two [001] planes of an fcc lattice, 
with each wall molecule attached to a lattice site by a harmonic spring. 
The mass of the wall molecule was set equal to that of the fluid molecule $m$. 
The number density of the wall was set at $\rho_w=1.86/\sigma^3$.
The wall-fluid interaction was modeled by another LJ potential $U_{wf}$
with the energy and range parameters given by 
$\epsilon_{wf}=1.16\epsilon$ and $\sigma_{wf}=1.04\sigma$, and $\delta_{wf}$ 
for specifying the wetting property of the fluid, taken to be $1$ 
in our simulations. The Couette flow was generated by moving 
the top and bottom walls at a constant speed $V_w$ in the $\pm x$ directions, 
respectively. In most of our simulations, the shearing speed was 
$V_w=0.05\sqrt{\epsilon/m}$, the sample dimension along $y$ was 
$6.8\sigma$, the wall separation along $z$ varied from $H=6.8\sigma$ 
to $68\sigma$, and the sample dimension along $x$ was set to be long 
enough so that the uniform single-phase shear flow was recovered 
far away from the MCL. Steady-state interfacial and velocity profiles 
were obtained from time average over $5\times 10^5\tau$ where $\tau$ 
is the atomic time scale $\sqrt{m\sigma^2/\epsilon}$.

The tangential slip velocity profiles next to the wall, i.e., 
the slip profiles, are shown in the inset to Fig. 2.  
While there is clearly a small core region, on the order of a few $l_s$, 
where the slip profiles display sharp decay, 
a much more gentle variation of the slip profiles becomes apparent 
as the system size $H$ increases. In order to quantify the nature of 
the gentle variation, we plot in Fig. 2 the same data in the log-log scale. 
The dashed line has a slope $-1$, indicating the $1/x$ behavior to be 
indeed realized in MD simulations. For our finite-sized systems, 
there is always a plateau in the slip velocity in each of the 
single-phase flows, also observable in MD simulations 
with a value given by $v_0^{slip}=2V_wl_s/(H+2l_s)$, 
which acts as an outer cutoff on the $1/x$ profile. 
This $v_0^{slip}$ expression is simply derived from 
the Navier-Stokes equation for uniform shear flow and the NBC.
From our largest MD simulation, the $v^{slip}\propto 1/x$ behavior 
is seen to extend to $\sim 50\sigma$
(or $\sim 25l_s$).  Hence as $H\rightarrow\infty$ and $v_0^{slip}$ 
approaches $0$ (no-slip), the power-law region can be 
very large indeed. A large $1/x$ partial slip region is significant, 
because the outer cutoff length scale directly determines 
the integrated effects, such as the total steady-state dissipation.  
While in the past the $1/x$ stress variation away from the MCL 
has been known \cite{hua}, to our knowledge the fact 
that the partial slip also exhibits the same spatial dependence 
has not been previously seen \cite{assumption},
even though the validity of the Navier boundary condition at 
high shear stress has been verified \cite{nbc,ckb}.

Since MD simulations reach size and accuracy 
(e.g., for $V_w/\sqrt{\epsilon/m}\ll 1$ ) limits quickly, 
a continuum hydrodynamic formulation is necessary for realistic simulations. 
In particular, continuum simulations are necessary for low flow rate 
immiscible flows, where MD simulations are known to be very resource intensive. 
Combining the GNBC with the Cahn-Hilliard (CH) hydrodynamic formulation 
of two-phase flow \cite{vinal,jacqmin}, 
we have obtained a continuum hydrodynamic model \cite{qws}, 
suitable for the calculation of much larger immiscible flows 
(than MD simulations) that are accurate to the molecular scale. 
In the same notations of Ref.\cite{qws}, the continuum model is 
formulated as follows. The two coupled equations of motion are 
the Navier-Stokes (NS) equation (with the addition of 
the capillary force density) and the CH convection-diffusion equation 
for the composition field $\phi({\bf r})=(\rho_2-\rho_1)/(\rho_2+\rho_1)$
(where $\rho_1$ and $\rho_2$ are the local number densities 
for the two fluid species):
\begin{mathletters}
\begin{equation}\label{he1}
m\rho\left[{\partial{\bf v}\over
\partial t}+ \left({\bf v}\cdot\nabla\right){\bf v} \right]=
-\nabla p +\nabla\cdot{\mbox{\boldmath$\sigma$}}^v
+\mu\nabla\phi+m\rho{\bf g}_{ext},
\end{equation}
\begin{equation}\label{he2}
{\partial\phi\over\partial t}+{\bf v}\cdot\nabla\phi=M\nabla^2\mu,
\end{equation}
together with the incompressibility condition 
$\nabla\cdot{\bf v}=0$. 
Here $m\rho$ is the average fluid mass density, $p$ is the pressure, 
${\mbox{\boldmath$\sigma$}}^v$ is the Newtonian viscous stress tensor, 
$\mu\nabla\phi$ is the capillary force density with 
$\mu=\delta F/\delta \phi$ being the chemical potential 
defined from the CH free energy functional $F$ \cite{free-energy},
$m\rho{\bf g}_{ext}$ is the external body
force density (for Poiseuille flows), and $M$ is the
phenomenological mobility coefficient. 
The boundary conditions at the solid surface are
$v_n=0$, $\partial_n \mu=0$ ($n$ denotes the outward surface normal),
the continuum form of the GNBC: 
\begin{equation}\label{he3} 
\beta v_x^{slip}=-\eta\partial_n v_x+
L(\phi)\partial_x\phi,
\end{equation}
and the relaxational equation for surface $\phi$:
\begin{equation}\label{he4} 
{\partial\phi\over\partial t}+{\bf v}\cdot\nabla\phi= 
-\Gamma  L(\phi), 
\end{equation}
\end{mathletters}
Here $L(\phi)=
K\partial_n\phi+\partial\gamma_{wf}(\phi)/\partial\phi$
with $\gamma_{wf}(\phi)$ being the wall-fluid interfacial free energy
density, $L(\phi)\partial_x\phi$ is the uncompensated Young stress,
and $\Gamma$ is a (positive) phenomenological parameter.

The continuum results shown in Fig. 2 were calculated on a uniform mesh, 
using the same set of material parameters and $M$, $\Gamma$ values
corresponding to the same local properties in all the five MD simulations. 
The overall agreement is excellent.
Such agreement is possible because the GNBC does not impose an artificial 
cutoff on the slip region. Below we extend the MCL simulations,
through continuum hydrodynamics, to lower flow rates and much larger systems.

The continuum simulation for macroscopic immiscible flows is a challenging task.
Methods based on a fixed uniform mesh would break down because 
it cannot afford to simulate macroscopic systems with molecular resolution 
near the MCL. We have employed for this problem the adaptive method
based on iterative grid redistribution \cite{ren-wang}.
The computational mesh is redistributed according to the
behavior of the continuum solution so that fine molecular
resolution is achieved in the interfacial region and near
the MCL, while elsewhere a much coarser mesh is used to save computational cost.
The mesh distribution is controlled by a monitor function 
(see \cite{ren-wang}), and the redistribution procedure is 
done repeatedly as the solution evolves to its steady state.  
A semi-implicit time stepping scheme is also used to speed up 
the approach to steady state.
For Couette flow, since the interface and contact line are confined 
in a region that is narrow in the $x$ direction but extended in 
the entire $z$ direction, we used a variable, adaptive grid 
in the $x$ direction and kept a uniform grid in the $z$ direction. 
This relatively simpler grid structure (compared to a general 
two dimensional variable grid)
greatly simplifies the space discretization as well as the matrix
structure in the implicit time discretization.

Figure 3 shows the continuum results for three large systems 
($H$ on the order of hundreds of $\sigma$) with small $V_w$ 
($\sim 0.001\sqrt{\epsilon/m}$, 
well beyond the accuracy of our MD simulations).
In all three cases, the capillary force was verified to be important 
only in the interfacial region. In fact it decays to zero 
(exponentially) within a few $\sigma$.
However, the pressure gradients and viscous forces 
shows a much slower variation. They are balanced outside 
the interfacial region, indicating the flow to be governed by the 
Stokes equation. This is expected, because the Reynolds number 
$m\rho V_wH/\eta
\approx 0.6$ for $\rho\approx 0.8/\sigma^3$, $V_w=0.005\sqrt{\epsilon/m}$,
$H\approx 300\sigma$, and $\eta\approx 2.0\sqrt{\epsilon m}/\sigma^2$.
In Fig. 3 the slip profiles, plotted on the log-log scale, clearly show the 
$1/x$ behavior extending from $|x|\approx 6l_s$ to 
$|x|\approx 270 l_s$. The inset to Fig. 3 shows the scaled tangential 
velocity profiles at the solid surface, from which the existence of 
universal slip profile is evident. Physically, when $H\gg l_s$, 
the regime of Stokes flow is governed by only one velocity scale $V_w$ 
and one length scale $l_s$.  Thus universality becomes evident 
in terms of $v_x/V_w$ plotted as a function of $x/l_s$.  
(The inset of Fig. 2 has already shown part of the universality,
that the core slip profile is independent of the system size $H$.)
We give a heuristic account of the universal slip profile as follows.
Away from the MCL, the viscous shear stress is given by 
$-a\eta v_x(x)/|x|$, where $a$ is a constant $\sim 1$, 
$\eta$ the viscosity, and $v_x(x)$ the local tangential velocity.
The NBC implies $v_x^{slip}(x)=-l_s a v_x(x)/|x|$. 
Since $v_x^{slip}(x)=v_x+V_w$,
combining the two equations yields 
$v_x^{slip}(x)/V_w=1/(1+{|x|}/{al_s})$ \cite{asymptotic}. 
This relation, with $a\approx 2.14$ for best fit, agrees with 
the continuum slip profiles extremely well, as seen in the inset to Fig. 3.

In the Couette geometry, external work is supplied to maintain 
the constant speed $V_w$ of the moving wall. 
The rate of work is given by the integral of the local 
tangential force $l_s^{-1}\eta|v^{slip}|$ times the wall speed $V_w$
\cite{qws}, i.e., 
$$\int \left( l_s^{-1}\eta |v^{slip}| \right)V_w dx 
=\eta V_w^2{\cal A}$$      
per unit length along $y$, where ${\cal A}$ 
is a numerical constant.
In the limit of $V_w\rightarrow 0$ and $l_s/H\rightarrow 0$,
${\cal A}$ is independent of $V_w$ and $l_s$ but depends on the outer 
cutoff of the $1/x$ profile. As the external work done in the steady state
must be fully dissipated, the total dissipation rate is equal to 
the rate of external work.

Slip profiles obtained from both MD and continuum simulations show 
that the $1/x$ partial-slip region starts from 
$x_c\approx 6l_s$ ($v^{slip}\approx 0.26V_w$).
The outer cutoff for the partial-slip region, denoted by $R$, 
is determined by the overall size of the system.
For the total dissipation, 
the contributions of the core region and the partial-slip region 
may be quantified by the dimensionless integrations
$${\cal A}={\cal C}=\int_0^{x_c}\left[\displaystyle\frac{|v^{slip}(x)|}{V_w}
\right]d(\displaystyle\frac{x}{l_s})
\approx 2.9$$
for the core and 
$${\cal A}={\cal P}=\int_{x_c}^R\left[\displaystyle\frac{|v^{slip}(x)|}{V_w} 
\right]d(\displaystyle\frac{x}{l_s})
\approx 2.14\ln\displaystyle\frac{2.14+R/l_s}{2.14+x_c/l_s},$$
for the $1/x$ region.
Assuming $l_s\sim 1\;$nm (a few $\sigma$'s) \cite{nbc}, 
for $x_c=10\;$nm and $R=1\;\mu$m we have ${\cal P}\approx 9.2$. While
for $x_c=10\;$nm and $R=1\;$mm we have ${\cal P}\approx 24$, i.e., 
the power-law region can contribute significantly more to 
the total dissipation than the core region.

The dissipation component that occurs at the 
fluid-solid interface can be evaluated as
$$\int \left( l_s^{-1}\eta |v^{slip}|^2 \right) dx  
=\eta V_w^2{\cal I}$$      
per unit length along $y$, where 
$${\cal I}=\int_0^{R}\left[\displaystyle\frac{|v^{slip}(x)|}{V_w} 
\right]^2d(\displaystyle\frac{x}{l_s})
\approx \displaystyle\frac{2.14R/l_s}{2.14+R/l_s}.$$
In the core region, the interfacial component of the dissipation is 
obtained by letting $R=x_c$, or 
${\cal I}\approx 1.6$, i.e., about $70\%$ of the total interfacial
dissipation (${\cal I}=2.14$ for $R/l_s\rightarrow \infty$).

Recently, there has been considerable interest in the transport 
at micro- and nanoscales. The lower limit for $R$ can reach 
submicrometer or even shorter length scales. 
On the other hand, there have been increasing evidences for large 
slip length realized in various fluid-solid interfaces 
\cite{barrat,tretheway,leger}. Slip length  $l_s$ as large as $1\;\mu$m 
has been reported \cite{tretheway,leger}. The results in this paper
indicate that fluid-solid interfacial dissipation is an important
contribution to the total dissipation if large slip length occurs
in a small system. While asymptotic analysis has shown that 
at large distances from the MCL, the flow field
is not sensitive to the slip boundary condition \cite{sensitivity},
yet the (macroscopic) asymptotic region may not be attained given 
the small system size and/or the large slip length. 
In this regard, a continuum hydrodynamic formulation of the contact-line 
motion is necessary for realistic simulations of fluid dynamics 
at micro- and nanoscales, as done in the present case.

Partial support from HKUST's EHIA funding and Hong Kong RGC
grants HKUST 6176/99P, 6143/01P, and 604803 is hereby acknowledged.

\begin{figure}
\centerline{\psfig{figure=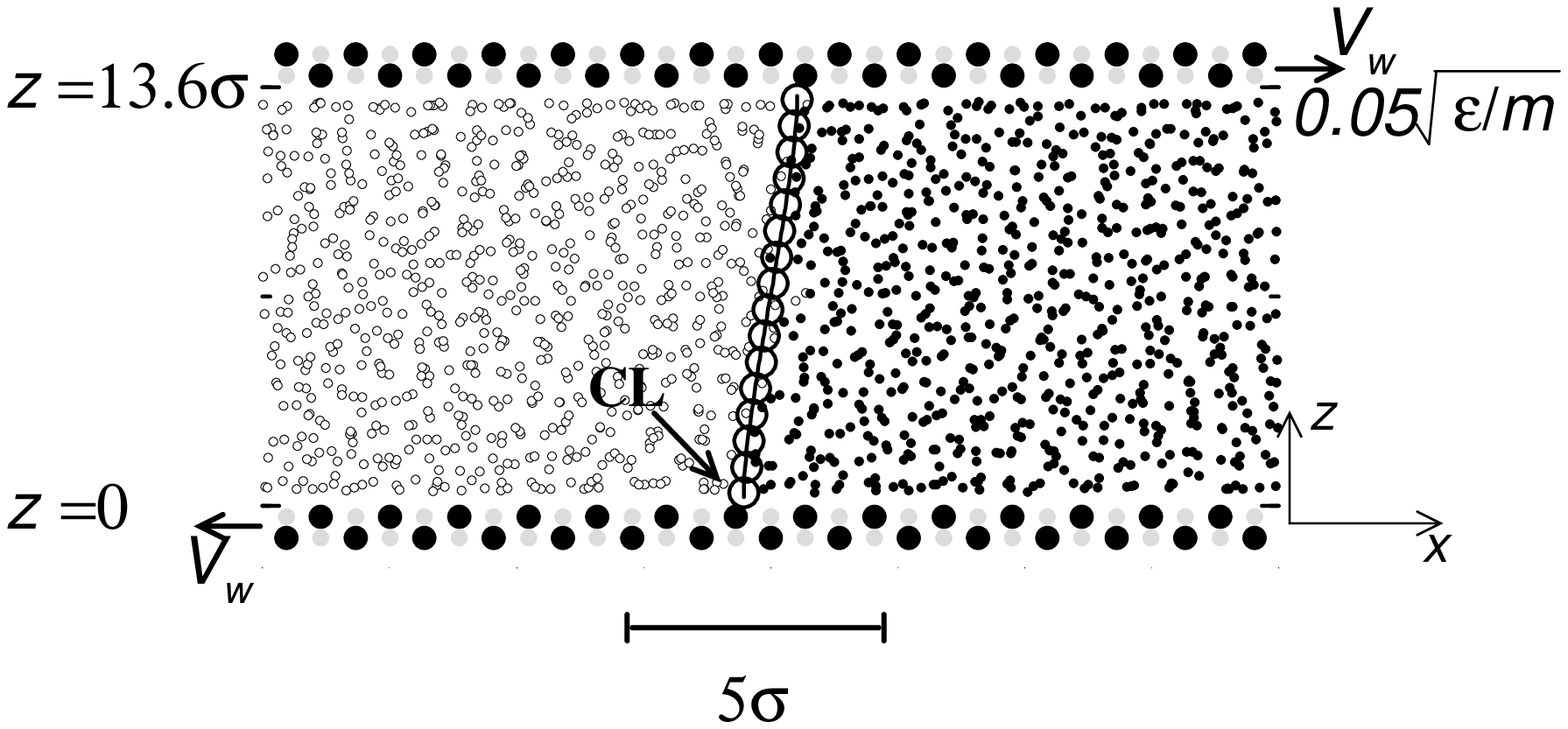,height=8.0cm}}
\bigskip
\caption{Segment of the MD simulation sample for an immiscible Couette 
flow with a $90^\circ$ static contact angle.
The small empty and solid circles indicate the instantaneous 
molecular positions of the two fluids projected onto the $xz$ plane. 
The horizontally aligned black/gray circles denote the wall molecules.
Between the two fluids, the large hollow circles represent the 
time-averaged interface profile, defined by $\rho_1=\rho_2$ ($\phi=0$). 
The solid line is the interface profile calculated from the continuum 
hydrodynamic model.}\label{fig1}
\end{figure}

\begin{figure}
\centerline{\psfig{figure=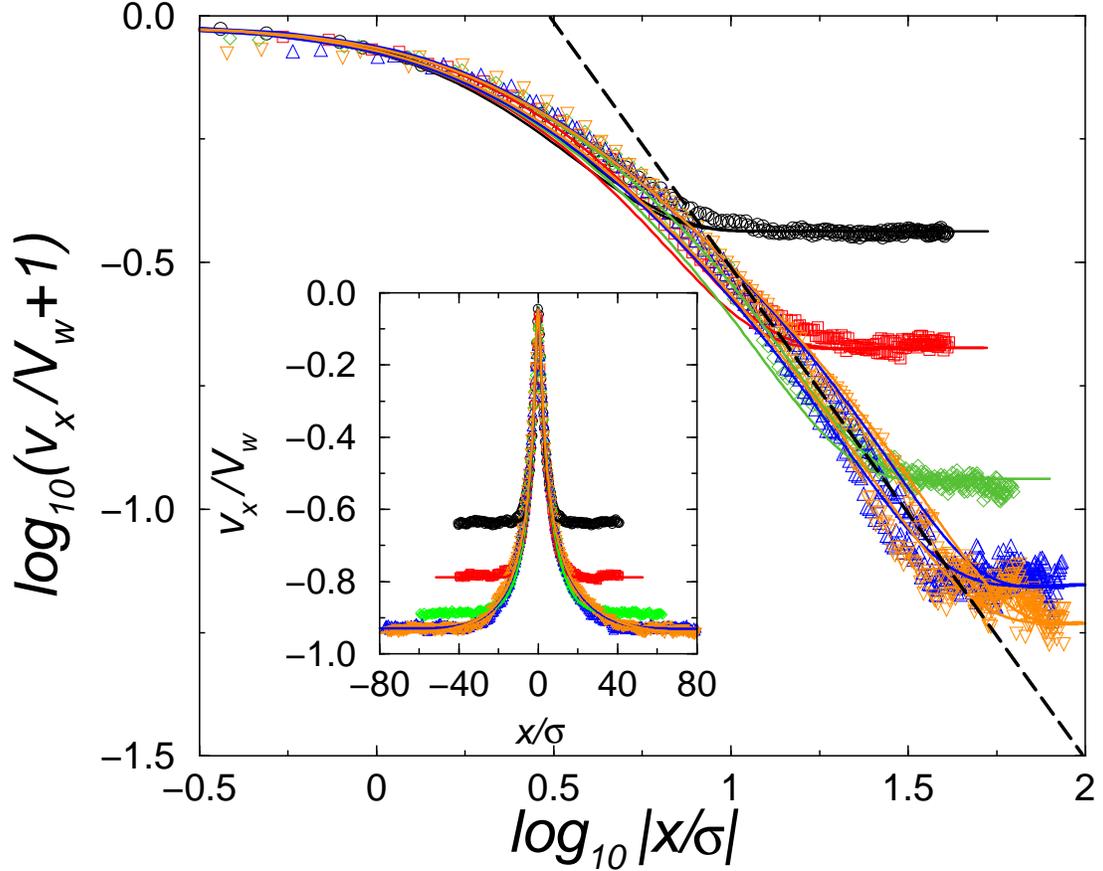,height=12.0cm}}
\caption{Log-log plot of the slip profiles showing the $1/x$ behavior. 
Here $v_x/V_w+1$ is the scaled slip velocity at the lower fluid-solid
interface $z=0$, and $x/\sigma$ measures the distance from the MCL 
in units of $\sigma$. The wall is moving at $-V_w$, hence $v_x/V_w=0$
means complete slip and $v_x/V_w=-1$ means no slip. 
The $v_x$ profiles were obtained for five symmetric cases of 
identical wall-fluid interactions for the two fluids 
(both $\delta_{wf}=1$ and thus a $90^\circ$ static contact angle). 
The five cases shown here used different values for $H$ but the same value 
for $V_w$ ($=0.05\sqrt{\epsilon/m}$) and also the same parameters 
for densities and interactions.
The symbols represent the MD results and the solid lines represent 
the continuum hydrodynamics results, obtained for 
$H=6.8\sigma$ (black circles and line), 
$H=13.6\sigma$ (red squares and line), 
$H=27.2\sigma$ (green diamonds and line), 
$H=54.4\sigma$ (blue up-triangles and line), 
$H=68\sigma$ (orange down-triangles and line).
The dashed line has the slope of $-1$, indicating that the $1/x$ behavior 
is approached for increasingly larger $H$. 
For $H=68\sigma$, the $1/x$ behavior extends from 
$|x|\approx 12\sigma\approx 6l_s$ to $50\sigma\approx 25l_s$, where
$l_s$ was measured to be $2\sigma$. 
Inset: The scaled tangential velocity $v_x/V_w$ at $z=0$
is plotted as a function of $x/\sigma$.}
\end{figure}

\begin{figure}
\centerline{\psfig{figure=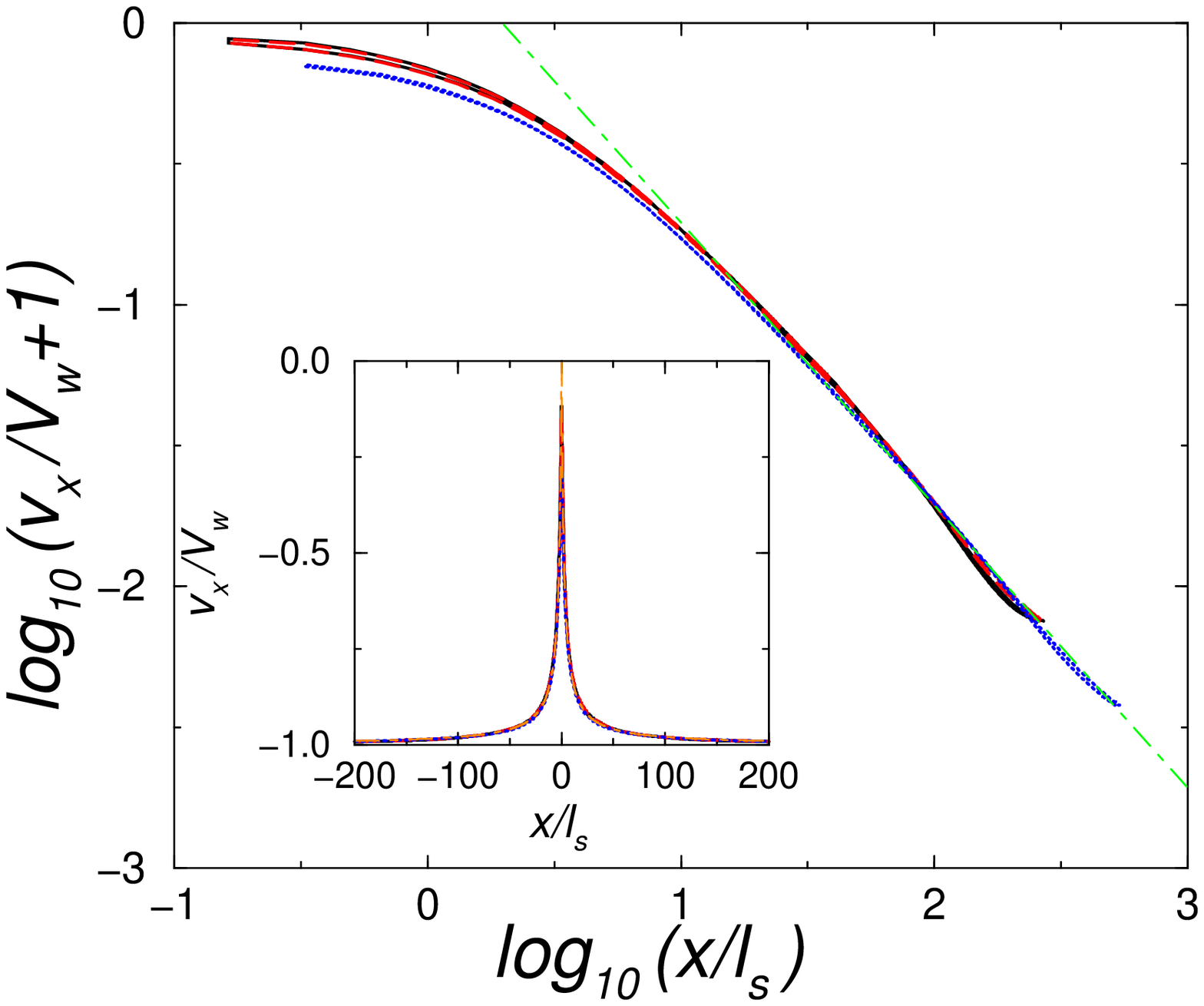,height=13.0cm}}
\caption{Log-log plot of the slip profiles showing the $1/x$ behavior. 
Here $v_x/V_w+1$ is the scaled slip velocity 
at the lower fluid-solid interface $z=0$,
and $x/l_s$ measures the distance from the MCL in unit of $l_s$. 
The wall is moving at $-V_w$, hence $v_x/V_w=0$ means complete
slip and $v_x/V_w=-1$ means no slip. 
The black solid line denotes the case of 
$H=326\sigma$, $V_w=0.005\sqrt{\epsilon/m}$ 
and $l_s=1.24\sigma$, the red dashed line denotes the case of
$H=326\sigma$, $V_w=0.0025\sqrt{\epsilon/m}$ and $l_s=1.24\sigma$, 
and the blue dotted line denotes the case of 
$H=326\sigma$, $V_w=0.0025\sqrt{\epsilon/m}$ and $l_s=0.62\sigma$.
The green dot-dashed line has the slope of $-1$, indicating a power-law region 
much wider than that in Fig. 2. Inset: Universal slip profile. 
The scaled tangential velocities $v_x/V_w$ at $z=0$ for all three cases
are plotted as a function of the scaled coordinate $x/l_s$. 
It is seen that the slip profiles 
show a partial-slip region as large as hundreds of $l_s$.
The relation $v_x/V_w=1/(1+{|x|}/{2.14l_s})-1$ 
is also plotted by the orange dot-dashed line, showing an extremely
good fit.}
\end{figure}


\begin{references}
\bibitem{assumption}
For example, an exponential slip profile was used in fitting the
MD results in Ref. \cite{robbins}.
\bibitem{asymptotic}
The similarity solution with a prescribed boundary velocity \cite{hua}
becomes the leading order solution in an asymptotic expansion 
in the slip length $l_s$, if the NBC is applied.
The leading order stress variation is $a\eta V_w/|x|$.
As a consequence, the leading order contribution to the slip velocity
is $al_sV_w/|x|$ (for small $l_s/|x|$).
\end{references}
\end{document}